%% file: main.tex
\documentclass[acmsmall,screen,noacm]{acmart}
\AtBeginDocument{%
  }

\setcopyright{none}
\renewcommand\footnotetextcopyrightpermission[1]{}

\input{packages}
\input{macro}

\begin{document}

\title{ReDON: \underline{Re}current \underline{D}iffractive \underline{O}ptical \underline{N}eural Processor with Reconfigurable Self-Modulated Nonlinearity}

\author{Ziang Yin}
\affiliation{%
  \institution{Arizona State University}
  \city{Tempe}
  \state{AZ}
  \country{USA}
}
\email{ziangyin@asu.edu}

\author{Qi Jing}
\affiliation{%
  \institution{Arizona State University}
  \city{Tempe}
  \state{AZ}
  \country{USA}
}
\email{qjing1@asu.edu}

\author{Raktim Sarma}
\affiliation{%
  \institution{Center for Integrated Nanotechnologies, Sandia National Laboratories}
  \city{Albuquerque}
  \state{NM}
  \country{USA}
}
\email{rsarma@sandia.gov}

\author{Zhaoran Rena Huang}
\affiliation{%
  \institution{Rensselaer Polytechnic Institute}
  \city{Troy}
  \state{NY}
  \country{USA}
}
\email{huangz3@rpi.edu}

\author{Yu Yao}
\affiliation{%
  \institution{Arizona State University}
  \city{Tempe}
  \state{AZ}
  \country{USA}
}
\email{yuyao@asu.edu}

\author{Jiaqi Gu}
\affiliation{%
  \institution{Arizona State University}
  \city{Tempe}
  \state{AZ}
  \country{USA}
}
\email{jiaqigu@asu.edu}

\input{doc/1_abs}

\maketitle
\input{doc/2_intro}

\input{doc/3_prelim}
\input{doc/4_algo}

\input{doc/5_exp}

\input{doc/6_conclu}

\bibliographystyle{ACM-Reference-Format}
\bibliography{ref/Top_sim,ref/Top, ref/addition,ref/Software,ref/NP,ref/ALG,ref/Cell,ref/PD,ref/DFM,ref/MPL,ref/NN, ref/IEEESettings}

\end{document}

%% file: packages.tex
\usepackage[utf8]{inputenc} %
\usepackage[T1]{fontenc}    %

\usepackage{booktabs}       %
\usepackage{soul}
\usepackage[table,xcdraw]{xcolor}
\usepackage{graphicx}
\usepackage{threeparttable}
\usepackage{multirow}
\usepackage{amsmath}
\usepackage{bm}
\usepackage{bbm}
\usepackage{amsthm}
\usepackage{enumitem} %
\usepackage[subrefformat=parens,labelformat=parens]{subfig}

\usepackage{wrapfig}
\usepackage[ruled, vlined, linesnumbered, noend]{algorithm2e}
\usepackage{amssymb} %

\SetCommentSty{mycommfont}

\SetKwComment{TriangleComment}{$\blacktriangleright$~}{} %
\usepackage{pifont}
\usepackage{bm}

%% file: macro.tex
\definecolor{citecolor}{RGB}{34,139,34}
\definecolor{mydarkblue}{rgb}{0,0.08,1}
\definecolor{mydarkgreen}{rgb}{0.02,0.6,0.02}
\definecolor{mydarkred}{rgb}{0.8,0.02,0.02}
\definecolor{mydarkorange}{rgb}{0.40,0.2,0.02}
\definecolor{mypurple}{RGB}{111,0,255}
\definecolor{myred}{rgb}{1.0,0.0,0.0}
\definecolor{mygold}{rgb}{0.75,0.6,0.12}
\definecolor{myblue}{rgb}{0,0.2,0.8}
\definecolor{mydarkgray}{rgb}{0.,0.2,0.2}

\definecolor{lightred}{RGB}{255,235,235}
\definecolor{lightgreen}{RGB}{235,255,235}
\definecolor{lightblue}{RGB}{235,235,255}
\definecolor{citelightblue}{RGB}{49,164,222}
\definecolor{lightcyan}{RGB}{235,255,255}
\definecolor{lightmagenta}{RGB}{255,235,255}
\definecolor{lightyellow}{RGB}{255,255,235}

\definecolor{qxkcolor}{RGB}{215,235,255}
\definecolor{softmaxcolor}{RGB}{230,235,255}
\definecolor{probxvcolor}{RGB}{255,255,235}

\definecolor{topkcolor}{RGB}{255,235,235}
\definecolor{zecolor}{RGB}{255,255,235}
\definecolor{dynacolor}{RGB}{235,255,255}

\definecolor{reviewcolor}{RGB}{0,0,200}

\theoremstyle{plain}

\theoremstyle{definition}

\newcommand{\titlename}{\texttt{ReDON}\xspace}

%% file: doc/1_abs.tex
\begin{abstract}
Diffractive optical neural networks (DONNs) have demonstrated unparalleled energy efficiency and parallelism by processing information directly in the optical domain. 
However, their computational expressivity is constrained by static, passive diffractive phase masks that lack efficient nonlinear responses and reprogrammability. 
To address these limitations, we introduce Recurrent Diffractive Optical Neural Processor (\titlename), a novel architecture featuring reconfigurable, recurrent self-modulated nonlinearity. 
This mechanism enables dynamic, input-dependent optical transmission through in-situ electro-optic self-modulation, providing a highly efficient and reprogrammable approach to optical computation. Inspired by the gated linear unit (GLU) in large language models, \titlename senses a fraction of the propagating optical field and modulates its phase or intensity via a lightweight, parametric function, enabling effective nonlinearity with minimal inference overhead.
As a non-von Neumann architecture with the main weighting units (metasurfaces) being fixed, we substantially extend the DONN's nonlinear representational capacity and task adaptability via recurrent optical hardware reuse and dynamically tunable nonlinearity.
We systematically investigate various self-modulation configurations to uncover the trade-offs between hardware efficiency and expressivity. 
On image recognition and segmentation tasks, \titlename improves test accuracy and mIoU by up to 20\% over prior DONNs with optical or digital nonlinearities at comparable complexity and negligible power overhead.
This work establishes a new paradigm for reconfigurable nonlinear optical computing, uniting the benefits of recurrence and self-modulation in non-von Neumann analog processors.

\end{abstract}

%% file: doc/2_intro.tex
\section{Introduction}
\label{sec:Intro}
Diffractive optical neural networks (DONNs) have emerged as a promising platform for high-throughput, low-power neuromorphic processing~\cite{NP_Nature2022_Luo, NP_DAC2025_Yin_CHORD, Lin2018AllOptical,Choi2025FreeSpaceEncoder,Majumdar2025MetaOpticalEncoders,Choi2025TransferableEncoder,Hu2024DiffractiveOpticalComputing}. 
By encoding neural weights directly into phase elements in the fabricated diffractive masks (e.g., metasurfaces), DONNs realize massively parallel in-memory computing, where weights are stored physically in subwavelength optical structures and applied to input with sub-nanosecond optical readout.
This non-von Neumann computational physical system, where computation and memory are co-located, has enabled ultra-efficient optical inference for applications including computer vision, sensing, holography, generative AI, and scientific computing~\cite{NP_Nature2022_Luo, NP_DAC2025_Yin_CHORD, Lin2018AllOptical,Tang2025OpticalPDE,Choi2025FreeSpaceEncoder}.

Despite these advantages, DONNs remain fundamentally limited by two key challenges: \textbf{weak optical nonlinearity and a lack of reconfigurability}, both rooted in the static, passive, and linear nature of diffractive metasurfaces.
\ding{202}~\textbf{Weak Nonlinearity}:~
Deep neural networks (DNNs) are fundamentally highly nonlinear functions for feature transformation. Yet DONNs are inherently (near) linear systems: cascaded passive diffractive layers reduce to a single global linear operator, with only weak square-law detection at photodetectors, dramatically restricting expressivity and limiting DONNs to tasks solvable by shallow NNs.
Researchers have explored several strategies to introduce nonlinearity.
Prior work has explored \emph{all-optical nonlinearity} (e.g., saturable absorbers, $\chi$(2)/$\chi$(3) nonlinear materials to build nonlinear optical encoder with enhanced accuracy, but often require unrealistically high optical power to trigger (100 kW/cm$^2$), have low energy efficiency (0.1\%), rooted in the shallow interaction depths (<1 $\mu m$) in free-space DONNs~\cite{Krasnok2017NonlinearMetasurfaces,Stich2025InverseDesign,Stich2025InverseDesignACSNano}.
\emph{Structure nonlinearity}, such as repeatedly encoding inputs~\cite{Yuan2023} or using cavity feedback~\cite{Xia2024}, provides high-order nonlinearity but lacks parametric control and acts more like fixed optical reservoirs than programmable activations.
To date, no solution offers efficient, programmable, low-latency nonlinear processing while maintaining DONNs’ speed and energy advantages.

\ding{203}~\textbf{Restricted Reconfigurability}:~
A second major limitation of DONNs is their \textbf{reconfigurability}.
Traditional DONNs depend on static metasurfaces whose nanostructures encode a fixed set of optical weight banks determined at fabrication time (similar to read-only memory). 
While this enables extremely fast, low-power inference (similar to memory readout), it prevents multi-channel, multi-layer neural computing and task adaptation in dynamic, evolving AI tasks.
Hybrid optical-electronic architectures treat DONNs as static optical encoders and rely on digital backends for task adaptation.
Multi-dimensional reconfiguration frameworks~\cite{NP_DAC2025_Yin_CHORD} and multi-task learning method~\cite{Zhou2025AutomatingMTL,Li2023RubiksONN} explore mechanical rotation, permutation, and wavelength/polarization tuning to introduce system-level reconfigurability, but mechanical actuation still shows limited reliability and speed.
As an active research topic, programmable metasurfaces can potentially offer per-element programmability but face substantial fabrication complexity, high insertion loss, and limited maturity for large-scale deployment~\cite{NP_NaturePhysRev2025_AbouHamdan}.
Overall, existing DONNs lack a scalable mechanism for input-dependent, high-expressivity computation without re-fabricating diffractive layers.
\input{figtex/fig_selfmodulation}

Inspired by the dynamic gating mechanisms of modern gated linear units (GLUs)~\cite{NP_arXiv2020_Shazeer}, we introduce \titlename, a diffractive optical neural processor that hybridizes fixed passive metasurfaces with a lightweight electro-optic self-modulation mechanism. 
\titlename senses a small fraction of the propagating optical field, processes it through a learnable parametric function, and uses the output to modulate the phase or amplitude of downstream diffractive layers. 
This introduces strong, tunable, input-dependent nonlinearity \emph{far beyond traditional pointwise}, static activations, like ReLU, while requiring negligible inference overhead.
Crucially, \titlename applies this modulation \textbf{recurrently} to reinforce nonlinearity and expand effective network depth.
This in-situ recurrence follows diffusion-style iterative refinement: the system composes multiple simple optical nonlinear steps into highly expressive mappings.
By hybridizing non-volatile metasurfaces with dynamic electro-optic self-modulation, \titlename overcomes long-standing limitations in nonlinearity and reconfigurability while preserving the efficiency and massive parallelism of diffractive optical in-memory computing.
Our main contributions are summarized as follows:
\begin{itemize}[leftmargin=*]
\setlength{\itemindent}{0.5em}
    \item \textbf{Tunable, Self-Modulated Electro-Optic Nonlinearity}:
    We introduce a novel diffractive optical nonlinear mechanism that senses intermediate optical fields and applies a learnable, GLU-inspired gating function to modulate downstream transmission, providing input-dependent, reprogrammable nonlinear transmission.
    \item \textbf{Recurrent Diffractive Optical Processing}:
    We introduce in-situ recurrence to reinforce the DONN nonlinearity by reusing the same hardware system with dynamic parameter tuning, incrementally composing deep neural representations for highly expressive transformations.
    \item \textbf{Hybrid Reconfigurable Optical Processor Architecture:}
    \titlename unifies static, non-volatile metasurface weight banks with lightweight electro-optic self-modulation to enable dynamic, reconfigurable computation in a non-von Neumann optical setting. 
    \item \textbf{Comprehensive Design Space Exploration}:
    We systematically analyze different architectural design settings and characterize their trade-offs in nonlinear expressivity, parameter efficiency, and hardware complexity. On classification and segmentation applications, our \titlename demonstrates an average of 20\% higher accuracy with superior task adaptability and negligible power overhead compared to existing linear/nonlinear DONNs.
\end{itemize}

%% file: figtex/fig_selfmodulation.tex
\begin{figure}
    \centering
    \includegraphics[width=0.85\columnwidth]{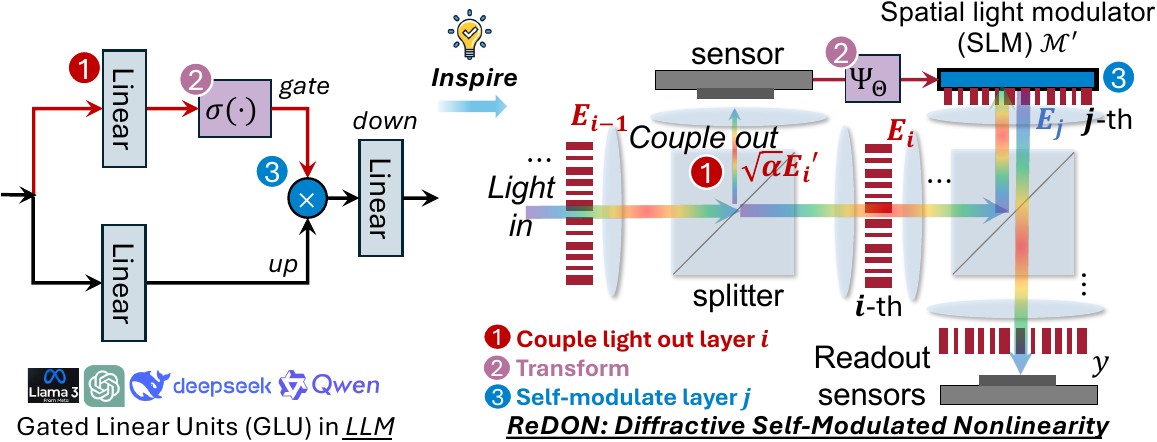}
    \vspace{-5pt}
    \caption{Inspired by GLU in LLMs, we propose a diffractive self-modulated nonlinear unit in \titlename. A small fraction of the light is sensed and self-modulates downstream metasurfaces.}
    \label{fig:SelfModulation}
    \vspace{-5pt}
\end{figure}

%% file: doc/3_prelim.tex
\section{Preliminary}
\label{sec:prelim}
A diffractive optical neural network (DONN)~\cite{NP_Nature2022_Luo, NP_DAC2025_Yin_CHORD, Lin2018AllOptical} implements computation through cascaded free-space propagation and metasurface modulation. Let $x_{\mathrm{in}}\in\mathbb{C}^N$ denote the input optical field. Each diffractive layer $i$ applies a diagonal transfer matrix $T_i=\mathrm{diag}(\exp(j\phi_i))$ representing spatially varying phase modulation, and propagation over distance $z_i$ is modeled by a linear operator $U_i(z_i)$ derived from the Fresnel diffraction integral. The end-to-end transformation is
\begin{equation}
    h_{\mathrm{out}}(x_{\mathrm{in}}) 
    = \left( \prod_{i=1}^{K} U_i(z_i)\, T_i \right) U_0\, x_{\mathrm{in}}, 
    \qquad 
    y = |h_{\mathrm{out}}|^2,
\end{equation}
where the square-law detection at the output plane provides the \emph{only} nonlinearity.  
Thus, once the metasurfaces are fabricated, the DONN implements a single fixed global linear operator in the field domain, followed by a static, uncontrollable intensity nonlinearity.  
The absence of tunable or distributed nonlinear mechanisms fundamentally limits expressivity and task adaptivity.

\noindent\textbf{Existing Optical Nonlinearity Mechanisms}.~
Recent work has explored several directions to introduce nonlinearity into diffractive or optical neural systems~\cite{Yildirim2024Nonlinear,Xia2025OnChip,Wang2023ImageSensing}.  
\emph{Structural nonlinearity} reuses the same linear diffractive elements multiple times so that the optical field is re-encoded and re-scattered through a static medium~\cite{Li2024}.  
Through repeated interaction and interference, the resulting intensity response effectively approximates a polynomial mapping~\cite{Xia2024}.  
These methods preserve low power and remain fully optical, but the nonlinear mapping is implicitly determined by the fixed geometry and cannot be shaped or tuned after fabrication.

\vspace{-5pt}
\subsection{Gated Linear Units}
Modern deep learning models, particularly large language models, increasingly rely on gated linear units (GLUs)~\cite{NP_arXiv2020_Shazeer} and related gating variants to achieve expressive, input-dependent transformations, e.g., SwigLU, SiLU. 
A GLU augments a linear projection with a learnable multiplicative gating function that modulates the output as
\begin{equation}
    \small
    \label{eq:GLU}
    z =GLU(x)= (W_{\text{up}}x+b_{\text{up}}) \odot \sigma\big((W_{\text{gate}}x+b_{\text{gate}})\big);~~y = (W_{\text{down}}z+b_{\text{down}}),
\end{equation}
where $\sigma(\cdot)$ is typically Swish or GELU.
The gating branch conditions the main signal, providing a flexible mechanism for selective amplification, suppression, and routing, offering dynamic modulation while maintaining computational and parameter efficiency.

%% file: doc/4_algo.tex
\vspace{-5pt}
\section{Recurrent Diffractive Optical Processor (\titlename)}
\label{sec:method}

We present \titlename, a recurrent diffractive optical neural processor. 
Unlike existing DONNs with fixed functionality and weak nonlinearity only at the photodetector arrays, \titlename introduces an electro-optic self-modulated nonlinearity and recurrent inference mechanism that simultaneously enables strong nonlinear expressivity and reconfigurability for non-von Neumann diffractive optical in-memory computing.
\input{figtex/fig_system_overview}
As summarized in Fig.~\ref{fig:SysmOverview}, \titlename admits a direct software-hardware correspondence: in software, a \titlename block wraps the optical operator $\mathcal{F}_{\titlename}(\cdot)$ with lightweight pointwise layers and a compact digital head, while in hardware the same computation is realized by sensing an intermediate optical field, computing a parametric transform $\Psi(\cdot,\Theta)$ on a lightweight electronic backend, and driving a modulation plane (e.g., SLM / tunable metasurface) to self-modulate downstream transmission. 
Recurrence reuses the same fabricated metasurface stack across iterations by updating only $\Theta$ while keeping $\Phi$ shared.

\vspace{-5pt}
\subsection{Diffractive Self-Modulated Nonlinearity}
\label{subsec:lc-phase}

Inspired by the powerful and parameter-efficient GLU mechanism in LLMs (Eq.~\eqref{eq:GLU}), we design \titlename with a diffractive self-modulation mechanism to introduce a dynamic input-dependent nonlinear response to DONNs.
As illustrated in Fig.~\ref{fig:SelfModulation} and Fig.~\ref{fig:SysmOverview}, we augment standard multi-layer DONNs with an auxiliary modulation branch that couples out and senses a small fraction ($\alpha$) of intermediate light field $E_i$ at the $i$-th metasurface, and convert it into electrical signals $\alpha|E_i|^2$.
In the electrical domain, we perform a lightweight parametric transformation function on the sensed signal $\Psi(E, \Theta)$ and use it to control the spatial-light modulator (SLMs) to modulate the phase or intensity for the main branch light field at subsequent metasurface layer(s).
As a simplified illustration, we assume the SLM modulates the $j$-th metasurface.
Let $E_i$ denote the complex amplitude of the optical field on the $i$-th metasurface output plane near-field, and $U_i(\lambda,z_i)$ denotes the linear diffraction matrix parameterized by wavelength $\lambda$ and diffraction distance $z_i$ between two metasurfaces. 
Each metasurface is a broadband phase mask that applies phase shifts $e^{j\Phi_i}$ to the incident light field.
The formulation of an $L$-layer \titlename system transmission is as follows
\begin{equation}
\small
\label{eq:titlename}
\begin{aligned}
    &\textcolor{gray}{\text{couple-out layer }i:}~~E_i'=U(z_i/2)E_{i-1},~E_{i-1}=\prod_{k=1}^{i-1}(e^{j\Phi_{k}}U_{k-1})E_0, E_0=\mathcal{E}(x)\\
    &\textcolor{gray}{\text{parametric transform:}}~~\mathcal{M}_j=\Psi(\alpha |E_i'|^2, \Theta),~~\mathcal{M}'=f(\mathcal{M})\\
    &\textcolor{gray}{\text{self-modulate layer}j:}~~E_j=\sqrt{1-\alpha}\mathcal{M}'_j\odot\prod_{k=i}^{j}(e^{j\Phi_{k}}U_{k-1})E_{i-1};\\
    &\textcolor{gray}{\text{transform after layer }j:}~~y=\mathcal{F}_{\titlename}(x)=(1-\alpha)\Big|U_L\prod_{k=j+1}^L(e^{j\Phi_k}U_{k-1})E_j\Big|^2;
\end{aligned}
\end{equation}
where the input $x\in[0,1]$ is first encoded as the incident light field $E_0$ via an input encoding function $\mathcal{E}(\cdot)$, e.g., phase encoding $E_0=e^{j\pi x}$, amplitude encoding $E_0=x$, etc. 
$\alpha|E_i'|^2$ is the intermediate light intensity coupled out between $i-1$ and $i$-th metasurfaces.
Here, $\Psi(\cdot,\Theta)$ generates control signals from the sensed light intensity, while $f(\cdot): \mathcal{M} \rightarrow \mathcal{M}'$ maps these signals to either phase shifts or attenuation factors at the modulated metasurface(s).”
For \textbf{phase self-modulation}, $\mathcal{M}'$ will be extra phase shifts within $[-\pi/2, \pi/2]$; for \textbf{intensity self-modulation}, $\mathcal{M}'$ will be mapped to real-valued attenuation factors.
\begin{equation}
    \small
    \label{eq:Modulation}
    \begin{aligned}
        \mathcal{M}'_{\text{phase}}&=f_{\text{phase}}(\mathcal{M})=e^{j(\mathcal{M} \% \pi - \pi/2)}\\
        \mathcal{M}'_{\text{intensity}}&=f_{\text{intensity}}(\mathcal{M})=\texttt{Clip}(\mathcal{M},~0,~ 1).
    \end{aligned}
\end{equation}

\input{figtex/fig_input_encoding}

\input{figtex/fig_nonlinear_fit}
\vspace{-5pt}
\subsection{Nonlinear Expressivity Investigation}
\noindent\textbf{Input Encoding Investigation}.~
The expressivity of the \titlename\ system is strongly influenced by the input encoding function $E_0=\mathcal{E}(x)$.
Figure~\ref{fig:InputEncoding} compares several commonly used encoding methods, including phase, amplitude, intensity, and complex-valued encoding.
Among them, \emph{phase encoding} consistently achieves the highest accuracy on classification tasks, whereas intensity and complex encodings exhibit noticeable training instability.
This observation aligns with prior findings~\cite{Li2024}, which show that phase encoding preserves more optical energy and introduces a natural nonlinear dependence in the photodetection process.
For these reasons, all subsequent experiments adopt the phase-encoding scheme $\mathcal{E}(x)=e^{j\pi x}, x\in[0,1]$.

\noindent\textbf{Scaled Differential Residual Output}.~
Since the output detection plane measures only optical \emph{intensity}, the resulting features are strictly non-negative, which limits representational richness and causes numerical instability during training.
A common workaround in prior optical neural networks is to use two parallel optical paths to form a differential signal, but this doubles the hardware cost.
To overcome the non-negative readout and avoid doubling hardware overhead, we instead form \emph{a scaled, differential residual path} between the input image and the detector readout, $y=x-\eta\mathcal{F}_{\titlename}(x)$.
The scaling factor $\eta$ controls the overall magnitude(norm) of the residual branch and compensates for the limited modulation range of phase-encoded diffractive propagation.
This formulation restores sign flexibility to the features while avoiding additional optical paths.

\noindent\textbf{Nonlinear Activation Expressivity}.~
The proposed self-modulation mechanism is fully learnable, enabling input-dependent nonlinear responses. 
To probe its expressivity, Fig.~\ref{fig:NonlinearFit} considers a minimal single-layer \titlename\ setup with $3\times 3$ input resolution and asks whether it can fit common nonlinear activation curves (compared to a linear DONN baseline). 
We synthesize controlled interference patterns by sweeping only the center input phase, $x\in[0,1]$ (field $e^{j\pi x}$), while keeping the remaining eight inputs at unit amplitude and zero phase. 
We then form a scalar output by reading the \emph{center} photodetector intensity (normalized), yielding a 1D mapping $x\mapsto y$; in contrast, aggregating all detector intensities is uninformative in our (approximately) lossless simulation because total optical power is conserved and thus nearly constant. 
By jointly optimizing the metasurface phases $\Phi$ and the coefficients $\Theta$ of a second-order polynomial modulation function $\Psi(\cdot,\Theta)$, \titlename\ accurately reproduces a variety of widely used nonlinear activations in Fig.~\ref{fig:NonlinearFit}. 
Existing linear DONNs, despite weak inherent nonlinearity, fail to approximate these functions and are constrained to nonnegative outputs under intensity-only readout. 
With the residual-output formulation, \titlename\ can realize sign-changing nonlinear mappings and, more generally, is not restricted to canonical pointwise activations; it can discover task-adaptive nonlinear functions during training.

\vspace{-5pt}
\subsection{Expressivity Augmentation with Multi-Layer Self-Modulation and Recurrent Architecture}

\noindent\textbf{Multi-Layer Self-Modulation}.~
In a standard GLU, only a single multiplicative gating operation is applied.
In a standard GLU, only a single multiplicative gating operation is applied.
To further strengthen the nonlinear capacity of our diffractive system, we extend this idea by allowing the same sensed intermediate signal to modulate \emph{multiple} downstream metasurface layers.
As illustrated in Fig.~\ref{fig:ParamSharing}(b), the optical field is coupled out at layer $i$, processed through individual $\Psi$ functions, and used to control any subsequent metasurface layer(s) $j$, where $j\ge i$.
For example, one may couple out at $i=1$ to modulate layers $j=2,3,5$.
Increasing the number of modulated layers naturally strengthens nonlinear expressivity but also incurs higher hardware cost.
In Sec.~\ref{sec:ModulationLayerExp}, we systematically evaluate this design space and quantify the trade-off between hardware complexity and expressivity gains achieved through multi-layer self-modulation.

\noindent\textbf{Recurrent Network Architecture}.~
Figure~\ref{fig:SysmOverview}(top) shows the overall model architecture using \titlename as the basic building block.
Lightweight pointwise convolutions are used for channel mixing and dimension matching, while the \titlename module performs per-channel spatial transformation.
Multiple \titlename blocks can be stacked to form a deeper backbone (with $N$ blocks), followed by a compact digital head for downstream tasks.
To further enhance nonlinear expressivity and maximize the utility of the non-volatile passive metasurfaces, we introduce a \textbf{recurrent inference mechanism}.
Each feature map is passed through the same \titlename system for $R$ iterations, effectively composing multiple nonlinear transformations in situ.
During recurrence, the metasurface phases $\Phi$ remain entirely shared and fixed, while we update per-channel $\Theta$ coefficients per recurrence iteration, which are stored in local SRAM and streamed to the modulation plane.
That is, \textbf{each hidden channel receives a distinct set of parameters at each recurrence iteration}, i.e., $\{\Theta^{r,c}\}, r\in [R], c\in [C_{in}]$.
With a fixed layer depth (i.e., constant $R\times N$), we find that \ul{recurrence of $R$=2 achieves the best expressivity and latency}, while more recurrence gives saturated gain.

\vspace{-5pt}
\subsection{Parameter-Efficient Coefficient Sharing}
The parametric function $\Psi$ introduces dynamically tunable weights $\Theta$, which must be stored in local SRAM and reloaded every inference. 
This incurs additional data movement, memory footprint, and compute overhead.
To reduce these costs while preserving high nonlinear expressivity, we explore two simple yet effective parameter-sharing strategies for $\Psi$.
Figure~\ref{fig:ParamSharing} summarizes the two approaches.

\noindent\textbf{Spatial Group-wise Coefficient Sharing}.~
Here, the $H\times W$ pixels on the modulation plane are divided into $p\times q$ spatial groups, each of size $r\times c$.
If $p=H, q=W$, each pixel has its own independent set of coefficients $\Theta$ in the $\Psi$ function, yielding maximal flexibility but the highest parameter count.
At the opposite extreme, when $p=q=1$, a single set of coefficients is shared across all modulated pixels.
By adjusting the group setting $p$ and $q$, we can smoothly trade expressivity for reduced memory and compute complexity, achieving up to $rc$ times reduction in parameters.

\noindent\textbf{Cross-layer Coefficient Sharing}.~ 
This strategy becomes relevant when the sensed signal from layer $i$ modulates multiple downstream metasurface layers $j$.
Instead of assigning distinct parameters $\Theta$ for multiple modulated layers, we can potentially share the $\Theta$ coefficients for all layers to reduce the parameter count and compute cost by $L$ times, where $L$ is the number of modulated layers.

\input{figtex/fig_param_sharing}

\vspace{-5pt}

\subsection{Implementation Practicality Justification}
\input{figtex/fig_two_impl}
Figure~\ref{fig:TwoImplementation} illustrates two feasible implementations of the proposed \titlename architecture.
In the reflective design (Fig.~\ref{fig:TwoImplementation}(a)), both the diffractive layers and light path are integrated within a single substrate. The back plane employs a gold reflector, while the intermediate optical field at layer $i$ is partially coupled out through a beam splitter and detected by an on-chip image sensor array.
A small number of high-speed ADCs can sequentially read out the sensed intensities in a time-multiplexed fashion, amortizing ADC cost.
The sensed signal $\alpha|E|^2$ can be transformed by $\Psi(\cdot,\Theta)$ in situ, either through compact analog circuits (for simple operations such as scaling, shifting, ReLU, or small-kernel convolutions) or lightweight digital units. The resulting control signal directly drives the modulation applied at the downstream metasurface layer(s) $j$.
Alternatively, a transmissive implementation (Fig.~\ref{fig:TwoImplementation}(b)) can be constructed by engineering metasurface $i$ to redirect a small fraction of the incident light ($\alpha$\%) toward an intermediate image sensor, while allowing the remaining light to propagate forward. 
Multiple metasurfaces can be fabricated on separate substrates with precise optical alignment.

\noindent\textbf{Modulation Plane}.~
A key hardware component is the tunable modulation plane at layer(s) $j$.
Commercial liquid-crystal spatial light modulators (LC-SLMs) already support reprogramming rates near 10 kHz, phase modulation exceeding $\pi$, and more than 8-bit precision, sufficient for many proof-of-concept real-time AI systems~\cite{Liu2023AOM_SLM}.
Looking forward, emerging technologies in actively programmable metasurfaces~\cite{NP_NaturePhysRev2025_AbouHamdan,NP_arXiv2025_Popescu}, promise significantly higher speeds and tighter integration.
Electro-optic modulation in particular can reach modulation rates in the \textbf{gigahertz range} with per-element tunability with compact footprints, which in principle, \emph{raise the achievable throughput by 2–3 orders of magnitude over LC-SLM-based prototypes}.
Indium-tin-oxide (ITO) transparent electrodes can be used to minimize optical loss.

%% file: figtex/fig_system_overview.tex
\begin{figure}
    \centering
    \includegraphics[width=0.85\columnwidth]{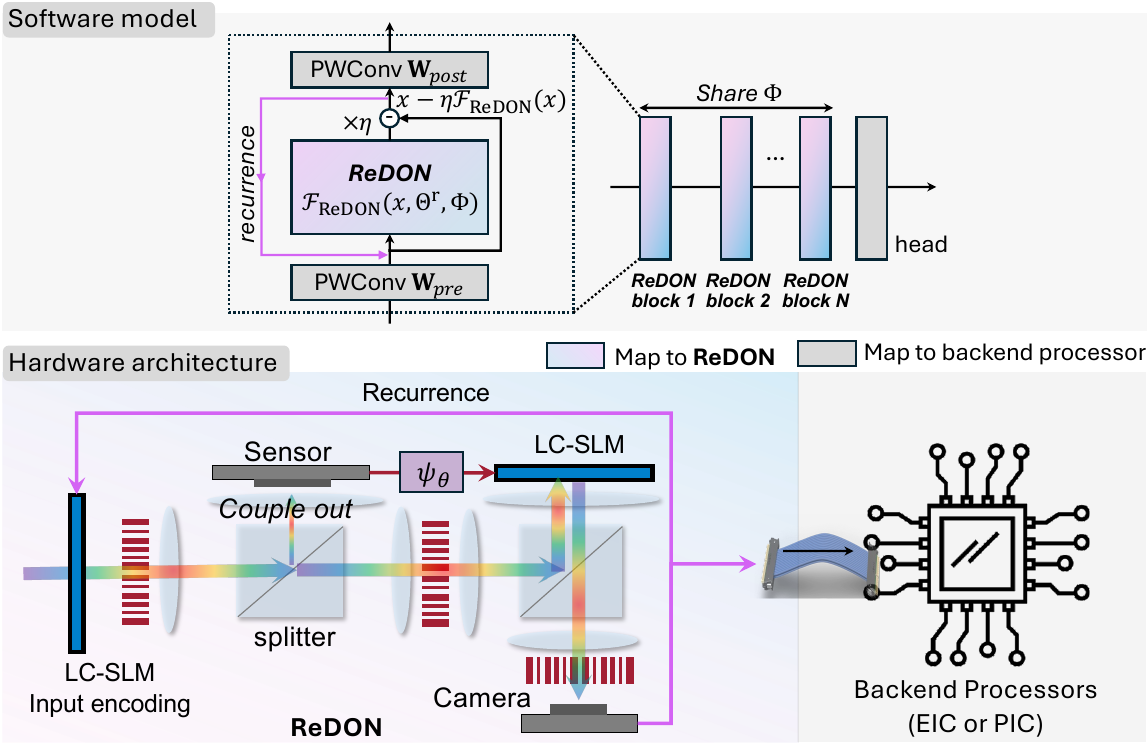}
    \vspace{-5pt}
    \caption{\titlename software--hardware correspondence. Top: software abstraction of a titlename block, where the optical operator $\mathcal{F}_{\text{titlename}}(\cdot)$ is integrated with lightweight pointwise layers and a compact digital head; recurrence reuses the same optical core with shared metasurface phases $\Phi$ and iteration-dependent modulation parameters $\Theta_r$. Bottom: a corresponding optoelectronic realization using an input encoder (e.g., LC-SLM), passive metasurface stack, intermediate optical sensing (coupling ratio $\alpha$), and a lightweight electronic processor that computes $\Psi(\cdot,\Theta)$ to drive a modulation plane for self-modulated nonlinearity.}
    \label{fig:SysmOverview}
    \vspace{-5pt}
\end{figure}

%% file: figtex/fig_input_encoding.tex
\begin{figure}
    \centering
    \includegraphics[width=0.55\columnwidth]{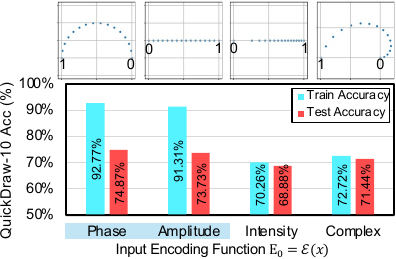}
    \vspace{-5pt}
    \caption{Compare different input encoding functions for \titlename on QuickDraw-10 classification. Phase and intensity encoding show the best effects.}
    \vspace{-5pt}
    \label{fig:InputEncoding}
\end{figure}

%% file: figtex/fig_nonlinear_fit.tex
\begin{figure}
    \centering
    \includegraphics[width=0.9\columnwidth]{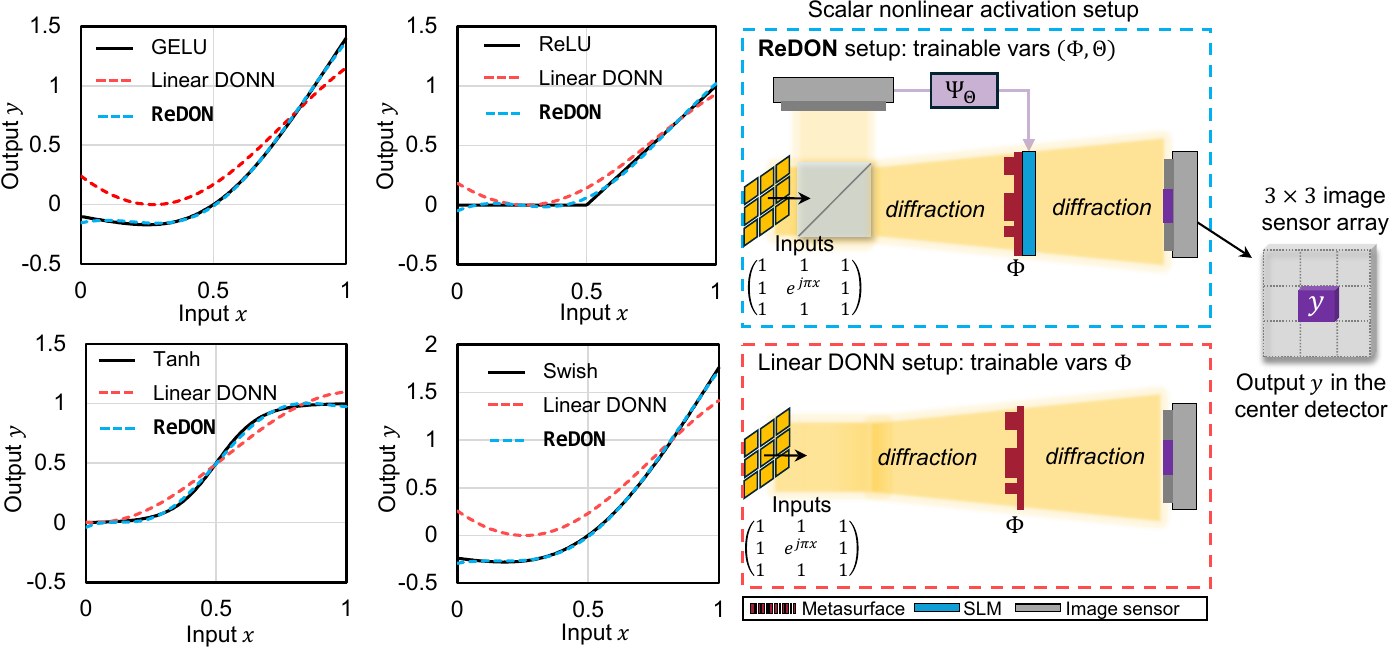}
    \vspace{-10pt}
    \caption{Evaluation of nonlinear expressivity by fitting popular activation functions. We simulate a $3\times 3$ phase-encoded input where only the center pixel sweeps phase $x\in[0,1]$ (others fixed at unit amplitude and zero phase) and read out only the center detector intensity to form a scalar mapping. ReDON (with a second-order polynomial $\Psi$ and residual readout $x-\eta\mathcal{F}_{\text{ReDON}}(x)$) approximates common nonlinear activations, while a linear DONN fails.}
    \label{fig:NonlinearFit}
    \vspace{-10pt}
\end{figure}

%% file: figtex/fig_param_sharing.tex
\begin{figure}
    \centering
    \includegraphics[width=0.9\columnwidth]{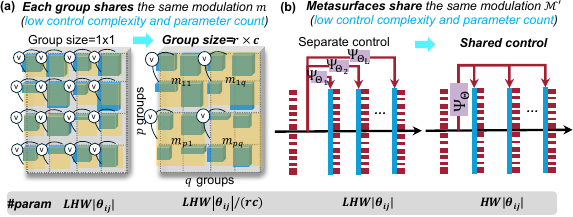}
    \vspace{-5pt}
    \caption{(a) Group-wise and (b) layer-wise parameter sharing on metasurface modulation.}
    \label{fig:ParamSharing}
    \vspace{-5pt}
\end{figure}

%% file: figtex/fig_two_impl.tex
\begin{figure}
    \centering
    \includegraphics[width=0.85\columnwidth]{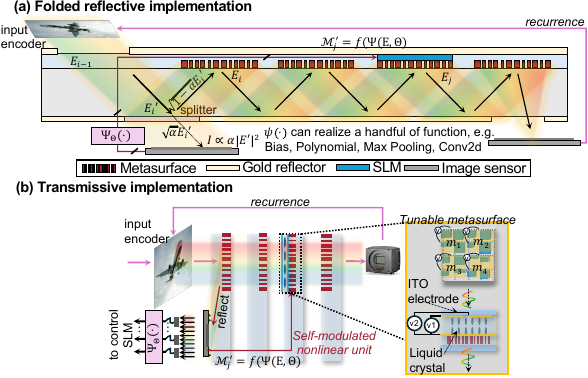}
    \vspace{-5pt}
    \caption{(a) Folded reflective and (b) transmissive example implementations of \titlename architecture. The self-modulation mechanism can be realized by image sensors and SLM.}
    \label{fig:TwoImplementation}
\end{figure}

%% file: doc/5_exp.tex
\vspace{-5pt}
\section{Evaluation}
\label{sec:evaluation}

\subsection{Evaluation Setup}
\subsubsection{Model and Dataset}
As a case study, we assume our \titlename system has up to 5 cascaded metasurfaces and 1 \titlename block.
We couple out $\alpha$=5\% of light.
The hidden dimension is 3.
For classification on CIFAR-10~\cite{NN_cifar2009} and QuickDraw-10/50~\cite{NN_QuickDraw2017}, the head is FC512ReLU-FC128ReLU-FC32ReLU-FC10.
For segmentation on binarized Stanford Background~\cite{5459211}, the head is Conv2DC128K3ReLU-Conv2DC64K3ReLU-Conv2DC2K1.
All 5 metasurfaces $\Phi$ \textbf{are shared} across all blocks.
Others parameters, including $\mathbf{W}_{pre}$, $\mathbf{W}_{post}$, $\eta$, $\Theta$ \textbf{are independently learned} for each \titlename block.
Each metasurface has 32$\times$32 meta-atoms. 
We use wavelength $\lambda = 532$ nm, meta-atom size $s = 400$ nm, and metasurface spacing $z = 8.42~\mu m$ as used in the literature~\cite{NP_Nature2022_Luo, NP_DAC2025_Yin_CHORD}. 

\vspace{-5pt}
\subsection{Ablation Study}
\subsubsection{Scaled, Differential Residual Output}
\input{figtex/fig_differential_output}
Figure~\ref{fig:DifferentialOutput} compares our proposed residual formulation $x-\eta\mathcal{F}(x)$ against several alternatives on the QuickDraw-10 classification task.
Both the explicit scaling factor and the differential residual path substantially improve expressivity and training stability.

\subsubsection{Parametric Function $\Psi$ Selection}
\input{figtex/fig_compare_nonlinear}
The design of the parametric function $\Psi$ plays a crucial role in balancing expressivity and hardware efficiency.
Figure~\ref{fig:CompareNonlinear} presents a comprehensive comparison across a wide range of candidates, including polynomials, common nonlinear operators (Tanh, ReLU, MaxPool), and lightweight operators such as 3x3 convolutions, evaluated under both phase and intensity self-modulation.
The functions are ordered by decreasing accuracy, revealing a trend consistent with their computational complexity.
Importantly, all candidates are implementable on compact digital logic, and several, such as scaling, bias, or small convolutions, can even be realized directly in the analog electrical domain without ADCs.
A key observation is that even the \emph{simplest affine transformations (scale + bias) significantly outperform the passive baseline (i.e., no self-modulation)}.
This demonstrates that \textbf{our self-modulation mechanism itself contributes the most to the expressivity, not relying on any complicated $\Psi$ function}.
For further evaluation, we select two nonlinear functions (third-order polynomial and scaled Tanh) and two linear functions (affine transform and simple bias) as representatives spanning a range of hardware costs.

\subsubsection{Modulation Layer Exploration}
\label{sec:ModulationLayerExp}
\input{figtex/fig_modulation_level}
We next examine how the choice of modulation layers affects expressivity and hardware cost.
Figure~\ref{fig:ModulationLevel} enumerates all combinations of coupling-out layers $i\in[1,5]$ and modulated layers $j\in[i,5]$.
For \textbf{phase self-modulation}, we observe two conclusions.
(1) \textbf{Later modulation layers yield higher accuracy}: For any fixed couple-out index $i$, accuracy increases as the modulation target $j$ moves deeper into the stack, with $j$=5 consistently offering the best performance.This suggests that a longer linear transformation path between layers $i$ and $j$ provides richer interference structure for modulation, whereas transformations occurring after modulation have comparatively limited impact.
(2) \textbf{Modulating more layers increases expressivity}:~
The strongest performance is achieved when all five layers are modulated, i.e., $i=1$ and $j=1,2,3,4,5$.

While for \textbf{intensity self-modulation}, however, increasing the number of modulated layers drastically reduces accuracy.
Unlike phase modulation, intensity modulation inherently attenuates optical power, and repeated attenuation across layers significantly limits expressivity due to compounded energy loss (and consequently low optical efficiency).
Based on these observations, we select representative configurations across different hardware budgets for later experiments: 3 to 5, 2 to (2,3), 1 to (2,4,5), 1 to (1,3,4,5), and 1 to (1,2,3,4,5).

\subsubsection{Parameter Sharing Strategy}
\label{sec:ParamShareExp}
\input{figtex/fig_group_size}
Figure~\ref{fig:GroupSize} evaluates different spatial parameter-sharing group sizes in the accuracy-parameter-count trade-off, using $i$=3, $j$=5 as a representative configuration.
Along the Pareto front, we identify four representative operating points, 32$\times$32, 8$\times$8, 32$\times$1, and 2$\times$2, that provide well-balanced trade-offs between expressivity and model compactness.
Depending on the available on-chip memory budget for storing the $\Theta$ parameters, one can choose an appropriate group size to meet system-level constraints.

\input{figtex/fig_param_reduction}
Figure~\ref{fig:ParameterReduction} further combines the optimal group-sharing configurations (from Fig.~\ref{fig:GroupSize}) with the best-performing modulation-layer settings (from Fig.~\ref{fig:ModulationLevel}) to reveal the overall accuracy-parameter-count landscape.
A clear trend emerges: \emph{modulating more layers has a stronger positive impact on expressivity than simply reducing the parameter-sharing group size}.
For instance, comparing 32$\times$32 3-to-5 vs. 16$\times$16 1-to-12345, with similar parameter counts, the latter consistently achieves higher accuracy.
It indicates that \textbf{modulating more diffractive layers contributes more to expressivity than increasing per-pixel parameter granularity}, suggesting a practical design rule for future nonlinear DONNs.

\vspace{-5pt}
\subsubsection{System Hardware Cost Analysis}
\noindent\textbf{\ul{Throughput Analysis}}.~
The primary throughput limitation of the proposed system stems from the modulation speed of commercially available spatial light modulators (SLMs), which currently operate at a maximum of $F_{\text{SLM,max}} \approx 10\,\text{kHz}$. 
Although our self-modulation mechanism substantially enhances nonlinear expressivity, its reliance on SLM reprogramming makes it the dominant inference speed bottleneck.
On the digital side, the detection head contains a total of 814{,}058 trainable parameters, corresponding to only $1.6\times10^6$ FLOPs, negligible compared with the DONN part.

To characterize the SLM modulation speed requirement, we model the required modulation frequency as
$F_{\text{SLM,req}} = R \cdot N \cdot C_{\text{mid}} \cdot \text{FPS}$,
where $R$ is the number of recurrent iterations, $N$ is the number of \titlename blocks, $C_{\text{mid}}$ denotes the hidden-channel dimension, and $\text{FPS}$ is the target application frame rate. 
The system, using a commercial liquid-crystal SLM, remains feasible when 
$F_{\text{SLM,req}} \leq F_{\text{SLM,max}} \approx 10\,\text{kHz}$.

Given the SLM speed limit and a spatial resolution of $32 \times 32$, the maximum digital throughput after photodetection and readout from the CMOS sensor is bounded by $f_{\text{sample}} = S \cdot F_{\text{SLM,req}},$
where $S$ is the total number of pixel readout samples. 
Under our typical configuration, this corresponds to 
$f_{\text{sample}} \approx 2.95 \ \text{MS/s},$
which is well within the capability of modern ADC/DAC hardware, commonly supporting sampling rates on the order of $10$--$14\,\text{GS/s}$~\cite{9731625, 9162776}.

Overall, with four \titlename blocks ($N$=4) and fewer than two recurrent iterations ($R$=2) per block, the system can process a 32$\times$32 spatial input with 3 hidden channels at up to 416 FPS, or 78 FPS when using 16 hidden channels.
These results demonstrate that \titlename remains compatible with real-time throughput.
As mentioned in the next section, one can potentially enable much higher throughput using a programmable metasurface that supports GHz modulation speed.

\noindent\textbf{\ul{Power Breakdown}}.~
We further analyze the power overhead related to the self-modulated circuitry.
The introduced power overhead is overall negligible ($<$1 mW) compared to the laser power ($>$100 mW) of the DONN system.
We estimate data-converter power using a standard figure-of-merit style scaling where power grows approximately linearly with sampling rate and exponentially with resolution (number of quantization levels). For a $b_{\text{in}}$-bit DAC and $b_{\text{out}}$-bit ADC operating at sampling rate $f$, we model:
\begin{equation}
    \small
    P_{DAC}(b_{in},f)=P_{0,DAC}\frac{2^{b_{in}}}{b_{in}+1}f,
\end{equation}
\begin{equation}
    \small
    P_{ADC}(b_{out},f)=P_{0,ADC}\frac{2^{b_{out}}}{b_{out}+1}f,
\end{equation}
where $P_{0,DAC}$ and $P_{0,ADC}$ are technology- and architecture-dependent constants derived from representative high-speed converter designs.

We estimate the power of the lightweight digital blocks (PWConv, $\Psi$ when implemented digitally, and the classification/segmentation head) by converting their required MAC throughput into power using an energy-per-MAC model. 
Using a 45\,nm INT8 MAC energy of $E_{\text{MAC}}\approx 0.23$\,pJ/MAC~\cite{HorowitzISSCC2014Energy}, the corresponding efficiency is $\eta_{\text{MAC}}=1/E_{\text{MAC}}\approx 4.35\times 10^{12}$ MAC/J. 
Therefore,
\begin{equation}
\small
P_{\text{digital}} = \frac{\text{MAC/s}}{\eta_{\text{MAC}}} = (\text{MAC/s})\cdot E_{\text{MAC}} .
\end{equation}

In Fig.~\ref{fig:PowerScale}, we show the power scaling with increased target FPS.
With a commercial liquid-crystal SLM that supports 10 kHz modulation rate, our \titlename can achieve over 400 FPS throughput.
A higher FPS can be realized using advanced electro-optic tunable metasurfaces~\cite{TI_DLPC410_DLPS024A_2012, becker2003mems, Benea-Chelmus2021}.
We can see SRAM (45 nm) access takes the largest proportion of power, while sensor array, TIA, ADC/DAC, and digital (PWConv/head) (45 nm), due to time-multiplexing and low cost, only take less than 3\% of the power overhead.
Our \titlename system ($N$=1) can potentially support over 1000 FPS end-to-end inference with only $\sim$1 mW electrical power overhead compared to passive DONNs, showing its potential for low-power edge deployment on real-time inference tasks.

\subsection{Main Results}
\label{sec:MainResult}
\input{tables/tab_main_results}
Using the optimal settings identified through our ablation studies, Table~\ref{tab:mainResult} compares our self-modulated nonlinearity with prior optical and hybrid nonlinear mechanisms.
We report both training accuracy (reflecting expressivity) and test accuracy (reflecting generalization).
For fully optical nonlinearities, we benchmark against saturable absorbers and reflection-based nonlinear coatings.
We also compare against natural nonlinearities arising from input encoding~\cite{Li2024}, as well as digital nonlinear activations (logarithmic, square, Tanh).
\textbf{All baselines use the same metasurface stack, input resolution, and digital head; only the nonlinearity mechanism differs}.
Across CIFAR-10, QuickDraw-50, and Stanford Background segmentation, our \titlename achieves substantially higher accuracy (\textbf{+20\%} on average) and mIoU, even with a single block ($N$=1) and a lightweight scaled Tanh $\Psi$ function.
Notably, prior DONNs with a single diffractive stage typically achieve $<$60\% accuracy on CIFAR-10 in the literature.
To our knowledge, \textbf{\titlename is the first to reach $\sim$65\% accuracy using only one block}, highlighting its markedly improved expressivity.

In the lower half of Table~\ref{tab:mainResult}, we perform a controlled ablation by keeping all architectural components identical but replacing our self-modulation mechanism with either (1) a purely linear DONN or (2) a DONN augmented with digital nonlinear activation.
Compared to simply adding digital activations, the self-modulation learns task-specific nonlinear functions with much higher accuracy, explaining the \textbf{clear gain from self-modulation over the digital activation function} at the same recurrence/block depth.
It also shows its \textbf{benefits compound when scaling to deeper architectures}.
Our evaluation focuses on modest-scale inputs for proof-of-concept evaluation, also due to training resource limitations.
Extending \titlename to high-resolution vision tasks is an important direction for future work.

\noindent\textbf{\underline{Task Adaptability of \titlename}}.~
\input{tables/tab_task_adaptation}
We assess \titlename’s task adaptability on both classification and PDE solving, as summarized in Table~\ref{tab:task_adapt}. 
After training on a source task, we \textbf{freeze all metasurfaces} and adapt only the head, PWConv, and the tunable $\Psi(\cdot,\Theta)$ functions on new tasks. 
Under this constraint, \titlename achieves substantially better transfer performance: on QuickDraw, it improves accuracy by 34\% over the baseline, and on Navier-Stokes it reduces the adapted solution error by 40\%.
We further compare \titlename to a nonlinear DONN based on saturable absorption~\cite{photonics12080779}, which can only finetune the head with a fixed diffractive optical encoder, which severely limits its transferability.

\noindent\textbf{\underline{\titlename Inference Visualization}}.~
\input{figtex/fig_visualization}
Figure~\ref{fig:Visualization} visualizes the inference results of \titlename on advanced learning tasks, including image segmentation and PDE solving, demonstrating the high expressivity and inference quality of \titlename in real-world complex tasks beyond simple classification.

\noindent\textbf{\underline{\titlename Robustness Study}}.~
\input{figtex/fig_robustness}
\input{tables/tab_robustness}
Practical diffractive optical systems inevitably suffer from physical non-idealities such as diffractive layer misalignment, random perturbations on the readout due to input/sensor noises, and fabrication-induced metasurface phase response deviations. 
We evaluate robustness on Fashion-MNIST classification using a 5-layer \titlename system \emph{without recurrence} to isolate the physical sensitivity of the optical front-end. 
As illustrated in Fig.~\ref{fig:Robustness}, we consider three non-ideality sources: (i) \emph{misalignment}, modeled as a random lateral shift of each diffractive plane, where the shift direction (horizontal or vertical) and sign are randomly selected, and the shift magnitude is set to the specified misalignment level (in pixel units with pixel size $s$=400nm); 
(ii) \emph{readout noise}, modeled as additive i.i.d. 
Gaussian noise $\mathcal{N}(0,\sigma^2)$ applied to the sensed intensities; 
and (iii) \emph{fabrication error}, modeled as i.i.d. Gaussian phase perturbations $\delta\phi\sim\mathcal{N}(0,\sigma^2)$ added to each meta-atom phase response.
For e-beam lithography fabricated metasurfaces, the average phase error experimentally observed is typically around 5 to 10 degrees, which justifies the practicality of phase error injection strength later in Table~\ref{tab:robustness}.

Table~\ref{tab:robustness} summarizes the resulting test accuracy under inference-time noise injection. 
For each noise setting in Table~\ref{tab:robustness}, we perform 5 independent evaluation iterations over the test set. 
Within an iteration, the injected noise realization is held fixed, while a new noise realization is resampled for the next iteration. 
Therefore, the evaluation is stochastic across iterations, and Table~\ref{tab:robustness} reports the mean and standard deviation of accuracy over 5 iterations.

Under standard (noise-free) training, \titlename remains relatively resilient to moderate MA and RN, but becomes increasingly sensitive to FE, and especially to compound system-level errors. In the most severe combined setting (MA 2 + RN 0.1 + FE 0.8), accuracy drops substantially. 
To mitigate this, we perform noise-aware training by injecting the combined system noise (MA 2 + RN 0.1 + FE 0.8) during training to encourage a more robust model.
This noise-aware model slightly reduces noise-free accuracy (92.9\%), but markedly improves robustness, maintaining 91-92\% accuracy across the tested ranges of MA/RN/FE and recovering most of the accuracy under harsh combined-noise conditions.

%% file: figtex/fig_differential_output.tex
\begin{figure}
    \centering
    \includegraphics[width=0.65\columnwidth]{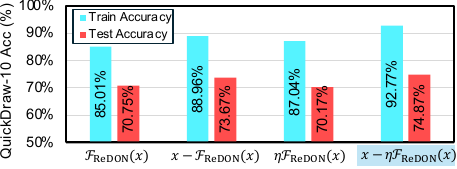}
    \vspace{-5pt}
    \caption{Compare different readout strategies. Our differential, scaled residual shows the best accuracy on QuickDraw-10.}
    \label{fig:DifferentialOutput}
    \vspace{-5pt}
\end{figure}

%% file: figtex/fig_compare_nonlinear.tex
\begin{figure}
    \centering
    \includegraphics[width=0.65\columnwidth]{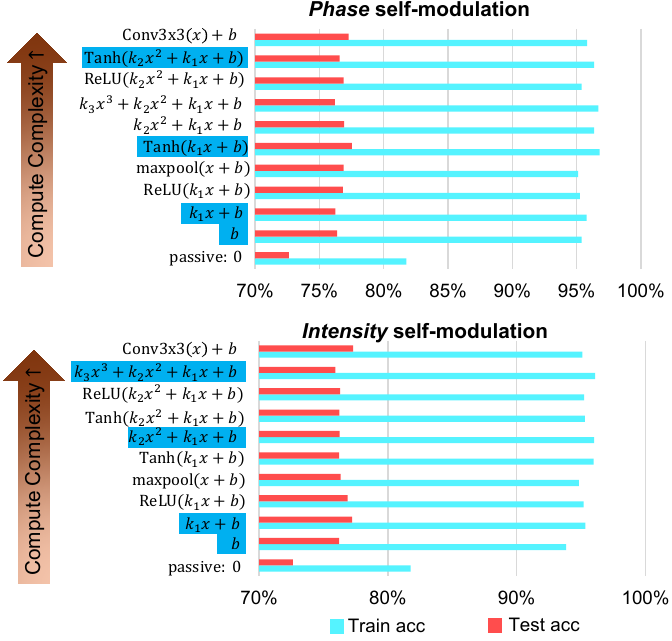}
    \vspace{-5pt}
    \caption{Compare the expressivity of different $\Psi(E,\Theta)$ functions on phase and intensity self-modulation. $\Psi(\cdot,\Theta)$ are ordered by computing complexity. Each highlighted function represents the highest-accuracy model at different complexity levels.}
    \label{fig:CompareNonlinear}
    \vspace{-5pt}
\end{figure}

%% file: figtex/fig_modulation_level.tex
\begin{figure*}
    \centering
    \includegraphics[width=\textwidth]{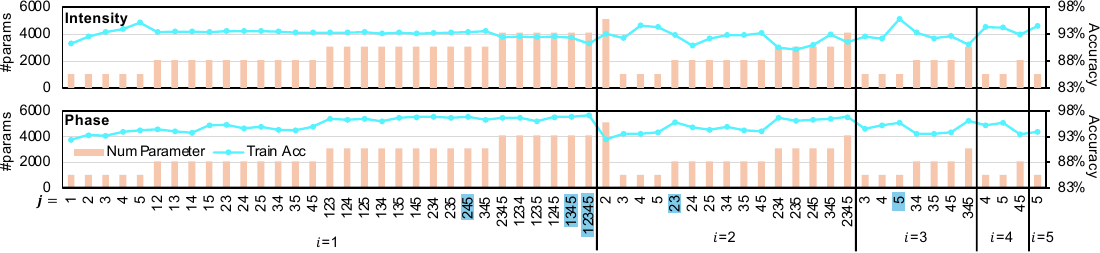}
    \vspace{-10pt}
    \caption{Parameter count and accuracy comparison with different sensing layer $i$ to modulated layers $j$ using phase and intensity self-modulation. Phase self-modulation shows better accuracy than intensity. The blue markup shows the best setting under different \#layers in $j$.
    These results motivate our choice of phase self-modulation and multi-layer modulation in the main experiments.}
    \label{fig:ModulationLevel}
    \vspace{-10pt}
\end{figure*}

%% file: figtex/fig_group_size.tex
\begin{figure}
    \centering
    \subfloat[]{\includegraphics[width=0.35\columnwidth]{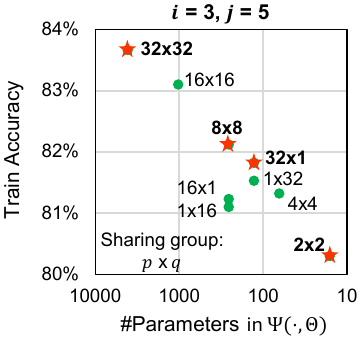}
    \label{fig:GroupSize}}
    \subfloat[]{\includegraphics[width=0.625\columnwidth]{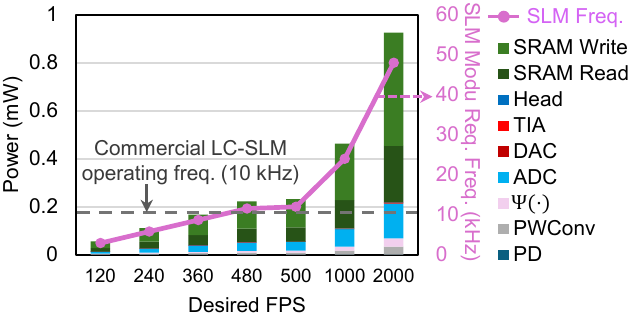}
    \label{fig:PowerScale}}
    \vspace{-5pt}
    \caption{(a) Accuracy and \#Params Pareto-front selection with different parameter sharing group sizes $p \times q$ on phase self-modulation with $i$=1, $j$=1,2,3,4,5 on QuickDraw-50. (b) Power breakdown and scaling with higher FPS.}
    \label{fig:GroupSize}
    \vspace{-3pt}
\end{figure}

%% file: figtex/fig_param_reduction.tex
\begin{figure}
    \centering
    \includegraphics[width=0.75\columnwidth]{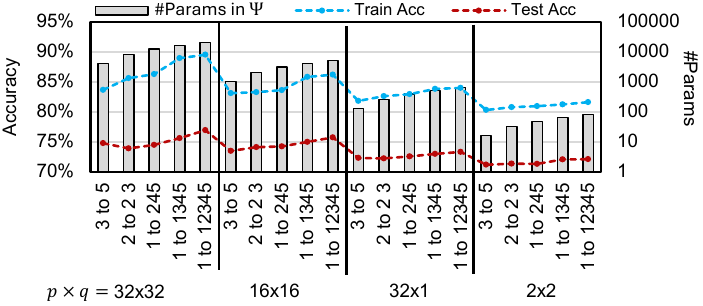}
    \vspace{-5pt}
    \caption{With the best modulation layer settings ($i$ to $j$) from Fig.~\ref{fig:ModulationLevel}, we explore \#Params vs. accuracy trade-off with different sharing group $p,q$.}
    \label{fig:ParameterReduction}
\end{figure}

%% file: tables/tab_main_results.tex
\begin{table}[]
\centering
\caption{Comparison across different nonlinearity designs on 3 benchmarks, our proposed \titlename with scaled Tanh as $\Psi$ consistently achieves the best performance.
\titlename shares $\Theta$ across all 5 modulated layers.
With more recurrence ($R$) and blocks ($N$), \titlename shows higher expressivity.
We replace our self-modulation with standard DONN or DONN with a digital Tanh activation function, which shows much worse results than our self-modulation.
}
\vspace{-10pt}
\resizebox{0.9\columnwidth}{!}{%
\begin{tabular}{c|cc|cc|cc}
\hline
                                                                                                                            & \multicolumn{2}{c|}{CIFAR-10} & \multicolumn{2}{c|}{QuickDraw-50} & \multicolumn{2}{c}{\begin{tabular}[c]{@{}c@{}}Stanford Background Seg.\\ Binarized Seg\end{tabular}} \\ \cline{2-7} 
\multirow{-2}{*}{Different Nonlinear Strategy}                                                                              & Train Acc.     & Test Acc.    & Train Acc.       & Test Acc.      & Train mIoU                                        & Test mIoU                                        \\ \hline
Saturable Absorption~\cite{photonics12080779}                                                                                                      & 61.9\%         & 61.1\%       & 80.1\%           & 71.3\%         & 72.2\%                                            & 47.1\%                                           \\ \hline
Reflection  Coating~\cite{dong2025scalablemultilayerdiffractiveneural}                                                                                                         & 58.4\%         & 58.3\%       & 76.6\%           & 68.5\%         & 68.3\%                                            & 41.3\%                                           \\ \hline
Encoding  (Phase)~\cite{Li2024}                                                                                                           & 61.2\%         & 60.2\%       & 79.4\%           & 70.4\%         & 76.1\%                                            & 53.4\%                                           \\ \hline
Encoding (Intensity)~\cite{Li2024}                                                                                                          & 58.3\%         & 58.6\%       & 76.5\%           & 68.7\%         & 66.4\%                                            & 46.9\%                                           \\ \hline
Encoding (Complex)~\cite{Li2024}                                                                                                            & 59.7\%         & 59.6\%       & 77.9\%           & 69.8\%         & 68.2\%                                            & 44.3\%                                           \\ \hline
Digital activation (Log)                                                                                                          & 60.1\%         & 58.1\%       & 78.3\%           & 68.3\%         & 69.5\%                                            & 38.2\%                                           \\ \hline
Digital activation (Square)                                                                                                       & 61.4\%         & 60.4\%       & 79.6\%           & 70.6\%         & 72.1\%                                            & 51.7\%                                           \\ \hline
Digital activation (Tanh)                                                                                                         & 62.6\%         & 60.7\%       & 80.7\%           & 70.9\%         & 77.2\%                                            & 55.6\%                                           \\ \hline
\rowcolor[HTML]{D3D3D3} 
\begin{tabular}[c]{@{}c@{}}\titlename $i$=1, $j$=1,2,3,4,5  \\ Tanh($k_1x+b$) cross-layer share\\ $p\times q$=32$\times$32, $R=1$, $N=1$\end{tabular} & 74.4\%         & 64.8\%       & 86.7\%           & 76.7\%         & 83.3\%                                            & 68.7\%                                           \\ \hline\hline
\begin{tabular}[c]{@{}c@{}}\titlename \\ Diffractive Only\\ $R=2$, $N=4$\end{tabular}                                         & 70.7\%         & 61.2\%       & 82.9\%           & 73.1\%         & 70.9\%                                            & 58.2\%                                           \\ \hline
\begin{tabular}[c]{@{}c@{}}\titlename \\ Digital Tanh \\ $R=2$, $N=4$\end{tabular}                                            & 77.4\%         & 66.4\%       & 91.8\%           & 77.6\%         & 85.4\%                                            & 63.8\%                                           \\ \hline
\rowcolor[HTML]{D3D3D3} 
\begin{tabular}[c]{@{}c@{}}\titlename $i$=1, $j$=1,2,3,4,5  \\ Tanh($k_1x+b$) cross-layer share\\ $p\times q$=32$\times$32, $R=2$, $N=4$\end{tabular} & 83.6\%         & 74.5\%       & 98.8\%           & 81.33\%        & 89.2\%                                            & 72.4\%                                           \\ \hline
\end{tabular}
}
\label{tab:mainResult}
\vspace{-5pt}
\end{table}

%% file: tables/tab_task_adaptation.tex
\begin{table}[]
\centering
\caption{Transferability evaluation of \titlename($R$=2, $B$=4).
On Fasion-MNIST$\rightarrow$QuickDraw-10, we show classification test accuracy.
On Darcy flow$\rightarrow$Navier-Stokes PDE solving, we show test mean-square error (MSE).
We train on the initial task and adapt to the second task, reusing fabricated metasurfaces.
}
\vspace{-5pt}
\resizebox{0.8\columnwidth}{!}{%
\begin{tabular}{c|cc|cc}
\hline
\multirow{2}{*}{Design} & \multicolumn{2}{c|}{Accuracy $\uparrow$}                                                                           & \multicolumn{2}{c}{Mean-Square Error (MSE) $\downarrow$}                                                                                  \\ \cline{2-5} 
                          & \multicolumn{1}{c|}{FMNIST} & \begin{tabular}[c]{@{}c@{}}QuickDraw-10 \\ adapt from FMNIST\end{tabular} & \multicolumn{1}{c|}{Darcy} & \begin{tabular}[c]{@{}c@{}}Navier-Stokes \\ adapted from Darcy\end{tabular} \\ \hline
Saturable Absorption~\cite{photonics12080779}     & \multicolumn{1}{c|}{80.1\%}  & 53.1\%                                                                    & \multicolumn{1}{c|}{0.095} & 0.1822                                                                      \\ \hline
\rowcolor[HTML]{D3D3D3} \titlename                     & \multicolumn{1}{c|}{95.1\%} & 87.2\%                                                                    & \multicolumn{1}{c|}{0.015} & 0.1035                                                                      \\ \hline
\end{tabular}
}
\label{tab:task_adapt}
\vspace{-5pt}
\end{table}

%% file: figtex/fig_visualization.tex
\begin{figure}
    \centering
    \includegraphics[width=0.6\linewidth]{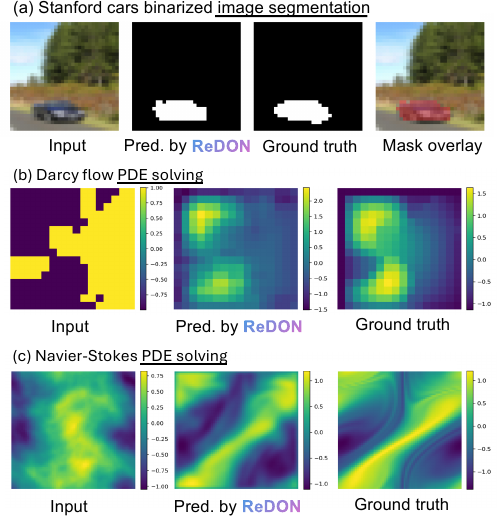}
    \vspace{-5pt}
    \caption{Visualization of representative inference results of \titlename on (a) Stanford cars image segmentation, (b) Darcy flow, and (c) Navier-Stokes PDE solving tasks.}
    \label{fig:Visualization}
    \vspace{-5pt}
\end{figure}

%% file: figtex/fig_robustness.tex
\begin{figure}
    \centering
    \includegraphics[width=0.76\columnwidth]{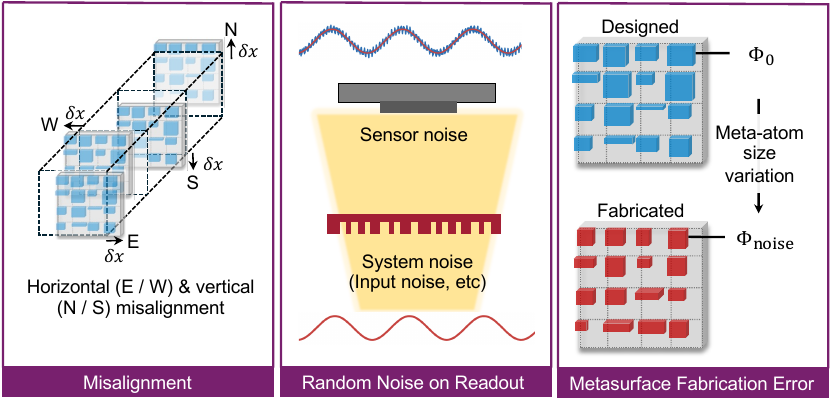}
    \vspace{-5pt}
    \caption{Robustness evaluation settings under practical non-idealities. 
    \emph{Left}: random horizontal/vertical misalignment of diffractive planes up to $\pm 1$ pixel/meta-atom size. 
    \emph{Middle}: additive Gaussian noise with $\sigma=0.01$ applied to sensed light intensities. 
    \emph{Right}: fabrication-induced phase error modeled as i.i.d. Gaussian phase perturbations with $\sigma_{\phi}=0.01~rad$ on metasurface phase responses.}
    \label{fig:Robustness}
    \vspace{-5pt}
\end{figure}

%% file: tables/tab_robustness.tex
\begin{table}[]
\centering
\caption{
Robustness of a 5-layer ReDON system (no recurrence) on Fashion-MNIST under misalignment (MA), readout noise (RN), fabrication error (FE), and their combination. 
The noise-free test accuracy is 95.1\%. 
The noise-aware model is trained with combined system errors (MA 2 + RN 0.1 + FE 0.8) injected during training; we report the average and std. of test accuracy over 5 iterations under inference-time noise injection.
}
\vspace{-5pt}
\resizebox{1\columnwidth}{!}{%

\begin{tabular}{cccccc}
\hline
\multicolumn{1}{c|}{\multirow{16}{*}{\begin{tabular}[c]{@{}c@{}}Noise-injected inference \\ (Standard Training)\end{tabular}}}                                       & \multicolumn{5}{c}{\cellcolor[HTML]{D3D3D3} Misalignment (MA)}                                                                                                                                                                                                   \\ \cline{2-6} 
\multicolumn{1}{c|}{}                                                                                                                            & \multicolumn{1}{c|}{Misaligned pixel size}                    & \multicolumn{1}{c|}{0.5}                       & \multicolumn{1}{c|}{1}                       & \multicolumn{1}{c|}{1.5}                       & 2                      \\ \cline{2-6} 
\multicolumn{1}{c|}{}                                                                                                                            & \multicolumn{1}{c|}{Averaged test accuracy}                   & \multicolumn{1}{c|}{92.6\%}                    & \multicolumn{1}{c|}{91.3\%}                  & \multicolumn{1}{c|}{90.1\%}                    & 88.2\%                 \\ \cline{2-6} 
\multicolumn{1}{c|}{}                                                                                                                            & \multicolumn{1}{c|}{Std. of test accuracy}                & \multicolumn{1}{c|}{0.0048}                    & \multicolumn{1}{c|}{0.0051}                  & \multicolumn{1}{c|}{0.0049}                    & 0.0053                 \\ \cline{2-6} 
\multicolumn{1}{c|}{}                                                                                                                            & \multicolumn{5}{c}{\cellcolor[HTML]{D3D3D3} Readout Noise (RN)}                                                                                                                                                                                                  \\ \cline{2-6} 
\multicolumn{1}{c|}{}                                                                                                                            & \multicolumn{1}{c|}{Injected Gaussian noise ($\sigma$)}       & \multicolumn{1}{c|}{0.04}                      & \multicolumn{1}{c|}{0.06}                    & \multicolumn{1}{c|}{0.08}                      & 0.1                    \\ \cline{2-6} 
\multicolumn{1}{c|}{}                                                                                                                            & \multicolumn{1}{c|}{Averaged test accuracy}                   & \multicolumn{1}{c|}{93.1\%}                    & \multicolumn{1}{c|}{92.4\%}                  & \multicolumn{1}{c|}{91.2\%}                    & 89.1\%                 \\ \cline{2-6} 
\multicolumn{1}{c|}{}                                                                                                                            & \multicolumn{1}{c|}{Std. of test accuracy}                & \multicolumn{1}{c|}{0.0029}                    & \multicolumn{1}{c|}{0.0028}                  & \multicolumn{1}{c|}{0.0029}                    & 0.0033                 \\ \cline{2-6} 
\multicolumn{1}{c|}{}                                                                                                                            & \multicolumn{5}{c}{\cellcolor[HTML]{D3D3D3} Fabrication Error (FE)}                                                                                                                                                                                              \\ \cline{2-6} 
\multicolumn{1}{c|}{}                                                                                                                            & \multicolumn{1}{c|}{Injected phase error $\sigma(\delta \phi)$} & \multicolumn{1}{c|}{0.2}                       & \multicolumn{1}{c|}{0.4}                     & \multicolumn{1}{c|}{0.6}                       & 0.8                    \\ \cline{2-6} 
\multicolumn{1}{c|}{}                                                                                                                            & \multicolumn{1}{c|}{Averaged test accuracy}                   & \multicolumn{1}{c|}{90.3\%}                    & \multicolumn{1}{c|}{88.2\%}                  & \multicolumn{1}{c|}{82.3\%}                    & 76.3\%                 \\ \cline{2-6} 
\multicolumn{1}{c|}{}                                                                                                                            & \multicolumn{1}{c|}{Std. of test accuracy}                & \multicolumn{1}{c|}{0.0046}                    & \multicolumn{1}{c|}{0.0052}                  & \multicolumn{1}{c|}{0.0054}                    & 0.0062                 \\ \cline{2-6} 
\multicolumn{1}{c|}{}                                                                                                                            & \multicolumn{5}{c}{\cellcolor[HTML]{D3D3D3} Combined: MA + RN + FE}                                                                                                                                                                                              \\ \cline{2-6} 
\multicolumn{1}{c|}{}                                                                                                                            & \multicolumn{1}{c|}{Combined system error}                    & \multicolumn{1}{c|}{MA 0.5 + RN 0.04 + FE 0.2} & \multicolumn{1}{c|}{MA 1 + RN 0.06 + FE 0.4} & \multicolumn{1}{c|}{MA 1.5 + RN 0.08 + FE 0.6} & MA 2 + RN 0.1 + FE 0.8 \\ \cline{2-6} 
\multicolumn{1}{c|}{}                                                                                                                            & \multicolumn{1}{c|}{Averaged test accuracy}                   & \multicolumn{1}{c|}{88.1\%}                    & \multicolumn{1}{c|}{86.2\%}                  & \multicolumn{1}{c|}{79.1\%}                    & 71.8\%                 \\ \cline{2-6} 
\multicolumn{1}{c|}{}                                                                                                                            & \multicolumn{1}{c|}{Std. of test accuracy}                & \multicolumn{1}{c|}{0.0052}                    & \multicolumn{1}{c|}{0.0054}                  & \multicolumn{1}{c|}{0.0051}                    & 0.0059                 \\ \hline
                                                                                                                                                 &                                                               &                                                &                                              &                                                &                        \\ \hline
\multicolumn{1}{c|}{\multirow{16}{*}{\begin{tabular}[c]{@{}c@{}}Noise-injected inference\\ (\textbf{Noise-aware Training})\end{tabular}}} & \multicolumn{5}{c}{\cellcolor[HTML]{D3D3D3} Misalignment (MA)}                                                                                                                                                                                                   \\ \cline{2-6} 
\multicolumn{1}{c|}{}                                                                                                                            & \multicolumn{1}{c|}{Misaligned pixel size}                    & \multicolumn{1}{c|}{0.5}                       & \multicolumn{1}{c|}{1}                       & \multicolumn{1}{c|}{1.5}                       & 2                      \\ \cline{2-6} 
\multicolumn{1}{c|}{}                                                                                                                            & \multicolumn{1}{c|}{Averaged test accuracy}                   & \multicolumn{1}{c|}{91.9\%}                    & \multicolumn{1}{c|}{91.7\%}                  & \multicolumn{1}{c|}{91.6\%}                    & 91.7\%                 \\ \cline{2-6} 
\multicolumn{1}{c|}{}                                                                                                                            & \multicolumn{1}{c|}{Std. of test accuracy}                & \multicolumn{1}{c|}{0.0041}                    & \multicolumn{1}{c|}{0.0049}                  & 0.0042                                         & 0.0044                 \\ \cline{2-6} 
\multicolumn{1}{c|}{}                                                                                                                            & \multicolumn{5}{c}{\cellcolor[HTML]{D3D3D3} Readout Noise (RN)}                                                                                                                                                                                                  \\ \cline{2-6} 
\multicolumn{1}{c|}{}                                                                                                                            & \multicolumn{1}{c|}{Injected Gaussian noise ($\sigma$)}       & \multicolumn{1}{c|}{0.04}                      & \multicolumn{1}{c|}{0.06}                    & \multicolumn{1}{c|}{0.08}                      & 0.1                    \\ \cline{2-6} 
\multicolumn{1}{c|}{}                                                                                                                            & \multicolumn{1}{c|}{Averaged test accuracy}                   & \multicolumn{1}{c|}{91.1\%}                    & \multicolumn{1}{c|}{91.0\%}                  & \multicolumn{1}{c|}{92.7\%}                    & 92.0\%                 \\ \cline{2-6} 
\multicolumn{1}{c|}{}                                                                                                                            & \multicolumn{1}{c|}{Std. of test accuracy}                & \multicolumn{1}{c|}{0.0028}                    & \multicolumn{1}{c|}{0.0029}                  & \multicolumn{1}{c|}{0.0031}                    & 0.0034                 \\ \cline{2-6} 
\multicolumn{1}{c|}{}                                                                                                                            & \multicolumn{5}{c}{\cellcolor[HTML]{D3D3D3} Fabrication Error (FE)}                                                                                                                                                                                              \\ \cline{2-6} 
\multicolumn{1}{c|}{}                                                                                                                            & \multicolumn{1}{c|}{Injected phase error $\sigma(\delta \phi)$} & \multicolumn{1}{c|}{0.2}                       & \multicolumn{1}{c|}{0.4}                     & \multicolumn{1}{c|}{0.6}                       & 0.8                    \\ \cline{2-6} 
\multicolumn{1}{c|}{}                                                                                                                            & \multicolumn{1}{c|}{Averaged test accuracy}                   & \multicolumn{1}{c|}{91.8\%}                    & \multicolumn{1}{c|}{91.2\%}                  & \multicolumn{1}{c|}{91.3\%}                    & 91.2\%                 \\ \cline{2-6} 
\multicolumn{1}{c|}{}                                                                                                                            & \multicolumn{1}{c|}{Std. of test accuracy}                & 0.0051                                         & 0.0057                                       & 0.0055                                         & 0.0048                 \\ \cline{2-6} 
\multicolumn{1}{c|}{}                                                                                                                            & \multicolumn{5}{c}{\cellcolor[HTML]{D3D3D3} Combined: MA + RN + FE}                                                                                                                                                                                              \\ \cline{2-6} 
\multicolumn{1}{c|}{}                                                                                                                            & \multicolumn{1}{c|}{Combined system error}                    & \multicolumn{1}{c|}{MA 0.5 + RN 0.04 + FE 0.2} & \multicolumn{1}{c|}{MA 1 + RN 0.06 + FE 0.4} & \multicolumn{1}{c|}{MA 1.5 + RN 0.08 + FE 0.6} & MA 2 + RN 0.1 + FE 0.8 \\ \cline{2-6} 
\multicolumn{1}{c|}{}                                                                                                                            & \multicolumn{1}{c|}{Averaged test accuracy}                   & \multicolumn{1}{c|}{91.6\%}                    & \multicolumn{1}{c|}{91.1\%}                  & \multicolumn{1}{c|}{91.5\%}                    & 91.1\%                 \\ \cline{2-6} 
\multicolumn{1}{c|}{}                                                                                                                            & \multicolumn{1}{c|}{Std. of test accuracy}                & \multicolumn{1}{c|}{0.0049}                    & \multicolumn{1}{c|}{0.0044}                  & \multicolumn{1}{c|}{0.0051}                    & 0.0052                 \\ \hline
\end{tabular}
}
\label{tab:robustness}
\vspace{-5pt}
\end{table}

%% file: doc/6_conclu.tex
\vspace{-5pt}
\section{Conclusion}
We present \titlename, a recurrent diffractive optical neural processor that overcomes the weak nonlinearity and rigidity of conventional DONNs through a self-modulated electro-optic nonlinearity. By sensing a small portion of the optical field and dynamically modulating downstream metasurfaces, \titlename provides a strong, tunable, input-dependent nonlinear response with minimal overhead while preserving the speed and parallelism of optical in-memory computing.
Through extensive design-space exploration, we show that \titlename consistently delivers far higher nonlinear expressivity with balanced efficiency compared to prior nonlinear mechanisms.
This makes \titlename a promising candidate for adaptive front-end encoders in optical machine learning systems.
It establishes a new direction toward reconfigurable, recurrent, nonlinear diffractive in-memory computing.

\begin{acks}
  R.Sarma and Y. Yao acknowledge support from the Laboratory Directed Research and Development program at Sandia National Laboratories and Sandia University Partnerships Network (SUPN) program. 
  This work was performed in part at the Center for Integrated Nanotechnologies, an Office of Science User Facility operated for the U.S. Department of Energy (DOE) Office of Science. 
  Sandia National Laboratories is a multimission laboratory managed and operated by National Technology \& Engineering Solutions of Sandia, LLC, a wholly owned subsidiary of Honeywell International, Inc., for the U.S. DOE's National Nuclear Security Administration under Contract No. DE-NA-0003525. 
  The views expressed in the article do not necessarily represent the views of the U.S. DOE or the United States Government.
\end{acks}